\documentclass[final,5p,times,twocolumn,letterpaper]{elsarticle}

\usepackage{natbib}
\usepackage{cancel}
\usepackage{ulem}
\usepackage{pifont}
\usepackage{amsthm}         
\usepackage{amsmath} 	   
\usepackage{amssymb}       
\usepackage{amsfonts}
\usepackage{graphicx} 	    
\usepackage{verbatim}
\usepackage{epsfig,bbm}
\usepackage{color}
\usepackage{colortbl}
\usepackage{float}
\usepackage{multirow}

\usepackage{subcaption}

\definecolor{babyblue}{rgb}{0.54, 0.81, 0.94}
\definecolor{corn}{rgb}{0.98, 0.93, 0.36}

\begin{document}

\begin{frontmatter}
\title{Fully stable cosmological solutions with a non-singular classical bounce}

\author[add0]{Anna Ijjas\corref{cor1}}
\ead{aijjas@princeton.edu}
\author[add0,add1]{Paul J. Steinhardt}
\address[add0]{Princeton Center for Theoretical Science, Princeton University, Princeton, NJ 08544 USA}
\address[add1]{Department of Physics, Princeton University, Princeton, NJ 08544, USA}

\cortext[cor1]{Corresponding author.}

\date{\today}

\begin{abstract}
We recently showed how it is possible to use a cubic Galileon action to construct classical cosmological solutions that enter a contracting null energy condition (NEC) violating phase, bounce  at finite values of the scale factor and exit into an expanding NEC-satisfying phase without encountering any singularities or pathologies.  A  drawback of these examples is that  singular behavior is encountered at some time either just before or just after the NEC-violating phase. In this Letter, we show that it is possible to circumvent this problem by extending our method to actions that include the next order ${\cal L}_4$ Galileon interaction.  Using this approach, we construct  non-singular classical bouncing cosmological solutions that are non-pathological for all times.
\end{abstract}


\end{frontmatter}

\section{Introduction} 

Cosmological scenarios that involve a phase of contraction followed by a bounce to a phase of expansion are of great interest since they can smooth and flatten the cosmological background \cite{Khoury:2001wf} and generate a nearly scale-invariant spectrum of super-horizon curvature modes \cite{Lehners:2007ac,Buchbinder:2007ad} while avoiding the multiverse and initial conditions problems of inflationary cosmology.   In these theories, the smoothing contraction phase is fully described by Einstein gravity and the quantum generation of curvature modes is described by standard semi-classical perturbation theory.  The challenge has been to find a non-pathological theoretical framework for describing the  bounce, {\it i.e.}, the transition from contraction to expansion. One possibility is to realize the transition through a singular (`quantum') bounce in which the scale factor passes through or tunnels through zero, as was proposed in \cite{Gielen:2015uaa};  this idea is intriguing, though the approach relies on some as-yet unproven assumptions about the analyticity of quantum gravity \cite{Bars:2013vba}.

Another approach is a `classical bounce,' in which the universe bounces after contracting to a small but finite size with energy density well below the Planck scale such that quantum gravity effects can be neglected.  The transition occurs through violation of the null energy condition (NEC) over a finite period of time that includes the bounce. On a smooth and flat Friedmann-Robertson-Walker (FRW)  cosmological background with $ds^2=-dt^2 + a^2(t)dx_idx^i$ (where $a(t)$ is the scale factor), NEC violation means that the Hubble parameter $H=\dot{a}/a$ increases with time, $\dot{H}>0$, where dot denotes differentiation with respect to time $t$.
The classical bounce has the advantage of not requiring any knowledge of quantum gravity. However, it is a well-known problem that NEC violation is prone to ghost or gradient instabilities or leads to singular behavior. 

In \cite{Ijjas:2016tpn}, we showed that it is possible to construct classical solutions that enter a contracting NEC-violating phase, bounce, and exit in an expanding NEC-satisfying phase without introducing pathologies by realizing the NEC-violating stage through a scalar field described by the generalized cubic Galileon action.  We presented an `inverse method' for constructing the solutions and used it to derive explicit examples.  The examples show that the universe can undergo a cosmological bounce at finite values of the scale factor and low energies well below the Planck scale without encountering singular behavior or requiring superluminal sound speed of co-moving curvature modes during the NEC-violating phase. 

A  feature of the examples based on the cubic Galileon action, though, is that singular behavior is always encountered at some time either shortly before or shortly after the NEC-violating phase.  Hence, the remaining open issue is whether these pathologies are inevitable or if a stable NEC-violating bounce stage can be embedded into a cosmology that is stable and non-singular throughout cosmic evolution. In this Letter, we show explicitly that it is possible to construct a fully stable bouncing cosmology by naturally extending our inverse method to actions that include the next order ${\cal L}_4$ Galileon interaction.
 
 This Letter is organized as follows: First, we give a brief review of Galileon cosmology and derive the stability conditions for linear-order scalar and tensor perturbations. Next, we explain why the cubic Galileon action inevitably leads to divergences and/or other singular behavior either just before or just after the stable NEC-violating phase in Sec.~\ref{singular}.  In Sec.~\ref{l4}, we show that, in principle, by extending the action to include the ${\cal L}_4$ Galileon interaction, the pathological behavior can be avoided for all times.  
In Sec.~\ref{examples}, we use our inverse method to construct explicit solutions.  (Readers only interested in the existence of non-pathological bouncing solutions may wish to jump to the figures.)  
 
 \section{Galileon cosmology}
 
We consider bouncing Galileon cosmologies (also known as Horndeski theories) described by the action of a single scalar field,
\begin{equation}
\label{action0}
S = \int d^4x \sqrt{-g}\,\left(\frac{1}{2} M_{\rm Pl}^2R + {\cal L}_{\phi} \right)\,,
\end{equation}
where $M_{\rm Pl}$ is the reduced Planck mass in the limit that the Galileon action converges to pure Einstein gravity (in our example, in the asymptotic past before or the asymptotic future after the bounce); $R$ is the Ricci scalar; and $g$ is the metric determinant. Throughout, we work in reduced Planck units ($M_{\rm Pl}^2=1$). The Lagrangian density of the Galileon field $\phi$ is given by 
\begin{equation}
\label{action01}
{\cal L}_{\phi} = \sum^5_{i=2}{\cal L}_i\,,
\end{equation}
with 
\begin{eqnarray}
\label{L0}
{\cal L}_2 &=& G_2(X,\phi),\quad \\
\label{L3}
{\cal L}_3 &=&  - G_3(X,\phi)\Box\phi ,\quad \\
\label{L4}
{\cal L}_4 &=& G_4(X,\phi)R + G_4,_{X}(X, \phi)\left( \left(\Box\phi\right)^2 - \left(\nabla_{\mu}\nabla_{\nu}\phi\right)^2 \right),\;\\
\label{L5}
{\cal L}_5 &=&  G_5(X, \phi)G_{\mu\nu}\nabla^{\mu}\nabla^{\nu}\phi +\\
&-& \frac{G_5,_{X}}{6}\left( \left(\Box\phi\right)^3 - 3\Box\phi \left(\nabla_{\mu}\nabla_{\nu}\phi\right)^2+ 2\left(\nabla_{\mu}\nabla_{\nu}\phi\right)^3\right) 
.\nonumber
\end{eqnarray}
Here,  $X=-(1/2)\nabla_{\mu}\phi\nabla^{\mu}\phi$ is the canonical quadratic kinetic term and the functions $G_i(X, \phi)$ characterize the $i$th Galileon interaction. The action was shown by Horndeski \cite{Horndeski:1974wa} to lead to classical equations of motion that are only second order in time derivatives, thus avoiding an Ostragadski ghost.  Horndeski and Galileon theories have had numerous applications in cosmology in the past, including massive gravity and attempts to explain late-time acceleration \cite{deRham:2012az}.  Here we apply them to construct stable cosmological solutions that violate the NEC and bounce.  

For the purposes of illustration, we shall work with the concrete example defined through the coupling functions
\begin{eqnarray}
\label{g2S}
G_2(X,\phi) &=& k(\phi)X + q(\phi)X^2 - V(\phi)\,,\\
G_3(X,\phi) &=& b(\phi)X\,,\\
G_4(X,\phi) &=&\frac{1}{2} f_1(\phi)+  f_2(\phi)X\,, \\
\label{g5S}
G_5(X,\phi) &=& 0
\,,
\end{eqnarray}
where $k(\phi)$ is the dimensionless quadratic coupling; $q(\phi)$ is the dimensionless quartic coupling;  $b(\phi)$ is the dimensionless coupling of the scalar field $\phi$ to the cubic Galileon term; and $V(\phi)$ is the scalar potential. The positive definite coupling to the Ricci scalar $(1/2)f_1(\phi)+f_2(\phi)X$ is distinctive of the ${\cal L}_4$ Galileon interaction. Note that positivity of the non-minimal coupling ensures that anti-gravity regions and associated instabilities are avoided.

 \subsection{Background}

On a spatially-flat, FRW cosmological background, the corresponding homogeneous equations of motion take the form
 \begin{eqnarray}
\label{EF1-FRW}
3 H^2  &=&  \frac{1}{2}k\dot{\phi}^2 +  \frac{1}{4}\left(3 q -2b'\right)\dot{\phi}^4  + V - 3f_1'H\dot{\phi} 
\quad
\\
&+& 3\left(b -  3f_2'\right)H\dot{\phi}^3  - 3\left(f_1-3f_2\dot{\phi}^2 \right)H^2 
\,,\quad \nonumber \\
\label{EF2-FRW}
-2\dot{H} &=&  \left(k+ f_1''\right)\dot{\phi}^2  +  \left(q-b'+f_2''\right) \dot{\phi}^4 + f_1'\ddot{\phi} 
\\
&+& \left(3f_2'- b\right)\ddot{\phi}\dot{\phi}^2
 - f_1'H\dot{\phi} + \left(3b-11f_2'\right)H \dot{\phi}^3 
\nonumber\\
&-&  4f_2H\dot{\phi}\ddot{\phi}   
+ 6f_2 \dot{\phi}^2H^2 +
 2\dot{H}\left( f_1 -f_2\dot{\phi}^2 \right) 
,\quad \nonumber
\end{eqnarray}
where prime denotes differentiation with respect to the scalar $\phi$. The first Friedmann equation characterizes the different contributions to the total energy density  $\rho_{\rm tot}$ while the second Friedmann equation gives the sum of total energy density and pressure $p_{\rm tot}$. The ratio $-\dot{H}/H^2$ is equal to the equation of state $\epsilon\equiv (3/2)(\rho_{\rm tot}+p_{\rm tot})/\rho_{\rm tot}$. It is a distinctive feature of the Galileon that there is a non-trivial mixing between spatial curvature terms and the scalar field, leading to possible ambiguities in defining the stress-energy tensor \cite{Khoury:2011da}. Here, we have followed the convention and derived $T_{\mu\nu}$  by varying the covariantized  theory with respect to the metric.
  
\subsection{Perturbations}

On a homogeneous background, the source of leading-order inhomogeneities is linear fluctuations of the metric and the scalar field. To ensure stability, the linear theory must not have any pathologies. 

To study the stability behavior of the linear theory, we expand the ADM decomposition of the Galileon action up to second order in perturbation theory.
The ADM metric is defined through 
\begin{equation}
ds^2 = -Ndt^2 + g_{ij}\left( N^idt + dx^i\right)\left( N^jdt + dx^j\right)
\,,
\end{equation}
where $N$ is the lapse, $N_i$ is the shift,  and $g_{ij}$ is the spatial metric. We parametrize linear-order perturbations to the shift and lapse as 
\begin{equation}
\delta N=N-\bar{N}=\alpha, \quad \delta N_i=N_i-\bar{N}_i=\partial_i\beta\,, 
\end{equation}
with $\bar{N}=1, \bar{N}_i= 0$ and $\bar{g}_{ij}=a^2 \delta_{ij}$ being the FRW background metric.
Since throughout we consider a single field, to study the perturbed scalar sector, we are free to choose the unitary gauge in which all spatial inhomogeneities are promoted to the  metric, 
\begin{equation}
g_{ij}=a^2 e^{2\zeta}\left(\delta_{ij}+h_{ij} +\frac{1}{2}h_{ik}h_{kj}\right)
,\end{equation}
while the scalar does not carry any perturbations, $\delta \phi\equiv0$. Here, $\zeta$ is the co-moving curvature mode and $h_{ij}$ is the linear-order tensor perturbation with $\partial_j h_{ij}=0, h^i_i=0$. Note that both $\zeta$ and $h_{ij}$ are gauge-invariant.

The second-order action for $\zeta, \alpha$ and $\beta$ is given by the Lagrangian
\begin{eqnarray}
\label{perturbed1}
{\cal L}^{(2)}_{\zeta, \alpha, \beta} &=& a^3 \Bigg(  -3A_h\dot{\zeta}^2+ B_h\frac{\left(\partial_i\zeta\right)^2}{a^2}
+ m_{\alpha}\alpha^2 
\\
&+& 6\gamma\alpha\dot{\zeta}   - 2 A_h\alpha\frac{\Delta\zeta}{a^2} + 2\frac{\Delta\beta}{a^2}\left( A_h\dot{\zeta} - \gamma\alpha \right)
\Bigg)\,,
\nonumber
\end{eqnarray}
where 
\begin{eqnarray}
\label{ah0}
A_{h}(t) &=& 1+f_1 - f_2\dot{\phi}^2 
\,,\\
\label{bh0}
B_{h}(t) &=& 1 + f_1+ f_2\dot{\phi}^2
\,,\\
\label{malpha}
m_{\alpha}(t) &=&  \frac{1}{2}k\dot{\phi}^2+\frac{1}{2}\left(3q-2b'
\right)\dot{\phi}^4  -3f_1'\dot{\phi} H  
\nonumber\\
&+& 6\left( b - 3f_2'  \right) H\dot{\phi}^3  
- 3\left( 1+f_1-6f_2\dot{\phi}^2 \right)H^2 
\,,\quad\\
\gamma(t) &=& \left( 1+f_1 - 3f_2\dot{\phi}^2 \right)H + \frac{1}{2}f_1'\dot{\phi} \\
&-& \frac{1}{2}\left( b - 3f_2' \right)\dot{\phi}^3 
. \qquad\nonumber
\end{eqnarray}
Varying the perturbed action with respect to the shift yields the momentum constraint which is, at the same time, a closed-form expression for the lapse $\alpha$,
\begin{equation}
\label{lapse}
A_h(t)\dot{\zeta} = \gamma(t)\alpha. 
\end{equation}
  The equation for the shift follows from substituting into the Hamiltonian constraint,
\begin{equation}
\label{shift}
\gamma^2(t)\frac{\Delta\beta}{a^2}=\left(m_{\alpha}(t)A_h(t)+3\gamma^2(t)\right)\dot{\zeta} -\gamma(t)A_h(t)\frac{\Delta\zeta}{a^2}. 
\end{equation}
Note that, for finite $\alpha$, if $\gamma=0$, $\dot{\zeta}=0$ or $A_h(t)=0$.

Substituting the expression for $\alpha$ and $\beta$ back into Eq.~\eqref{perturbed1}, we obtain the second-order action for co-moving curvature modes $\zeta$,
\begin{equation}\label{secondzeta}
S^{(2)}_{\zeta}=\int d^4x a^3 \left( A(t)\dot{\zeta}^2- B(t)\frac{\left(\partial_i\zeta\right)^2}{a^2}\right)\,,
\end{equation}
where the coefficients of the kinetic and gradient terms are defined as
\begin{eqnarray}
\label{Aeq}
A(t) &=&m_{\alpha}(t) \left(\frac{A_h(t)}{\gamma(t)}\right)^2 + 3A_h(t)
\,,\\
\label{Beq}
B(t) &=& a^{-1}(t)\frac{d}{dt}\left(a(t)\frac{A_h^2(t)}{\gamma(t)}\right) - B_h(t)
\,.
\end{eqnarray}
Similarly, the second-order action for tensor modes takes the simple form
\begin{equation}
S^{(2)}_{h_{ij}}=\int d^4x a^3 \left( A_h(t)\dot{h}^2_{ij}- B_h(t)\frac{\left(\partial_l h_{ij}\right)^2}{a^2}\right)
\,,
\end{equation}
where the coefficients of the kinetic and gradient terms are as defined above in Eqs.~(\ref{ah0}-\ref{bh0}). (The perturbed Horndeski action was previously obtained in \cite{Kobayashi:2011nu}; here we re-derived the action for our specific example to emphasize some pedagogical points, in particular, the role of the quantity we call $\gamma$.)

For the theory to be linearly stable, both the scalar and the tensor sector have to be stable, {\it i.e.}, the coefficients $A, B, A_h,$ or $B_h$ must not become negative.

\section{Singular behavior with ${\cal L}_3$ only} 
\label{singular}

In Ref.~\cite{Ijjas:2016tpn}, we showed that it is possible with the cubic (${\cal L}_3$) Galileon action ({\it i.e.}, without including ${\cal L}_4$ and ${\cal L}_5$)  to enter a NEC-violating phase, bounce, and restore the NEC without encountering any instability or singularity during this bounce phase.  However, there remained bad behavior either shortly before or shortly after the NEC-violating phase, depending on whether $\gamma$ was positive or negative when entering the NEC-violating phase.  

To see the problem, note that Eq.~(\ref{Beq}),  which determines gradient stability or instability, can be rearranged and integrated from $t$ to $t_0$, where $t_0$ is the time at the beginning of the NEC-violating phase and $t<t_0$, to yield the relation
\begin{equation}\label{BeqInt}
\frac{a A_h^2}{\gamma}\bigg|_{t}=\frac{a A_h^2}{\gamma}\bigg|_{t_0} - \int_{t}^{t_0} a(B+B_h) dt \equiv \frac{a A_h^2}{\gamma}\bigg|_{t_0} - I(t)\,.
\end{equation}
Without loss of generality, we take $\gamma(t_0)=\gamma_0$ positive.  Hence, the first term on the right hand side is non-negative.

In the case of the cubic Galileon action, $A_h=B_h=1$.    For a non-singular bounce where $a>0$ for all $t$, the integrand in the second term is positive definite, so $I(t)$ is divergent as $t \rightarrow -\infty$.  Consequently, $a A_h^2/\gamma$ must become negative at some finite $t<t_0$, which is where trouble is encountered.  Either $\gamma$ diverges (which requires couplings and/or kinetic energies to diverge) or $\gamma$ passes through zero, which requires that $a$ or $B$ in the integrand  become negative and diverge.  As emphasized in \cite{Ijjas:2016tpn}, this trouble occurs outside the NEC-violating regime, more precisely at $t<t_0$, and is not directly related to NEC or the bounce; it is a feature of the cubic Galileon action.

\section{Removing the singular behavior with ${\cal L}_4$ }
\label{l4}
The introduction of the next order Galileon interaction, ${\cal L}_4$, provides the freedom for $A_h$ and $B_h$ to become functions of time, which makes it possible to satisfy Eq.~(\ref{BeqInt}) for all times without encountering any bad behavior.  
  
First, we show that it is possible to have smooth evolution if $\gamma(t)$ crosses continuously through zero at some finite time $t_{\gamma}$ (in particular, $\gamma(t) \sim \gamma_0 (t- t_\gamma)$ at zero-crossing with $\gamma_0 >0$) if we choose $A_h$ and $B_h$  to be proportional to $(t-t_{\gamma})^2$. Because $t_\gamma$ can be well before NEC violation, we have the freedom to choose $A_h(t)$ and $B_h(t)$  to approach unity throughout the bounce phase.  This means that \ ${\cal L}_4$ and ${\cal L}_5$  are negligible during the bounce phase, and we automatically recover the same stable, non-singular behavior  during the bounce described in \cite{Ijjas:2016tpn}  assuming the ${\cal L}_3$ action only.  We also have the freedom to choose $A_h(t)$ and $B_h(t)$ to approach unity and the coefficient of the ${\cal L}_3$ term to approach zero in the asymptotic past and future, which ensures that the past converges to a contracting universe described by Einstein gravity and the future converges to an expanding universe described by Einstein gravity.  That is, it is straightforward to design models in which the Galileon is only active for a short period surrounding the bounce and pure Einstein gravity applies at all other times.
Notably, since the scale factor $a(t)$ diverges in the infinite past and infinite future, the total solution with bounce is geodesically complete.  

As illustrated in the example below, we can simplify further by choosing $A_h(t)=B_h(t)$, so that the sound speed for tensor perturbations is 
\begin{equation}
c_T^2=\frac{B_h}{A_h}=1 \quad \textnormal{for all}\;t\,.
\end{equation}
Near the zero-crossing point $t_{\gamma}$,   $A_h(t)= B_h(t) \sim A_h^0(t-t_{\gamma})^2$. Since, for all finite times $t$, both $a(t)$ and $H(t)$ are finite, it is straightforward to determine the behavior of $A, B$ at any time $t$. In particular, in the vicinity of the zero-crossing point,  $
\gamma \sim \gamma_0(t-t_{\gamma})$ where $\gamma_0>0$ ,  $a \rightarrow  a(t_{\gamma})>0$; $H \rightarrow H(t_{\gamma})$; and 
\begin{eqnarray}
\label{A-gamma-crossing}
A &\sim& A_h^0\left(m_{\alpha}(t_{\gamma})\left(\frac{A_h^0}{\gamma_0}\right)^2+3\right)(t-t_{\gamma})^2 + {\cal O}((t-t_{\gamma})^4)  
\,, \\
\label{B-gamma-crossing}
B&\sim& A_h^0\left(3\frac{A_h^0}{\gamma_0}-1\right)(t-t_{\gamma})^2 + {\cal O}((t-t_{\gamma})^4).
\end{eqnarray}
Finally, the expression for the sound speed of co-moving curvature modes simplifies to 
\begin{equation}
\label{cS-gamma-crossing}
c_S^2 \to  \frac{3\frac{A_h^0}{\gamma_0}-1}{m_{\alpha}(t_{\gamma})\left(\frac{A_h^0}{\gamma_0}\right)^2+3} \quad {\rm as}\; t\to t_{\gamma}
\,.
\end{equation}
As we we will illustrate in the next section by giving a concrete example (see Fig.~\ref{fig3}), we have the freedom to choose the coefficients $A_h^0$ and $\gamma_0$ such that $c_S^2 \sim {\cal O}(1)$ for all times.

If the second order action Eq.~(\ref{secondzeta}) is re-expressed in terms of the canonical variable $v\equiv z\zeta$ (with $z^2=a^2(t)A(t)$), then the equation of motion for $v$,
\begin{equation}
\label{v-eq}
\ddot{v} + \left(c_S^2k^2 - \frac{\ddot{z}}{z}\right)v = 0\,,
\end{equation}
has the solution $v=\, {\rm\it constant}$ near $t_\gamma$, where the constant is exponentially small if the preceding contraction consists of an ekpyrotic smoothing phase. The same conclusion can be drawn for the canonically normalized tensor modes  $u= z_h h_{ij}$,
\begin{equation}
\label{u-eq}
\ddot{u} + \left(c_T^2k^2 - \frac{\ddot{z}_h}{z_h}\right)u = 0\,,
\end{equation}
where $z_h^2= a^2(t)A_h(t)$ and $c_T^2\equiv1$.
(Technically, the derivatives in Eq.~\eqref{v-eq}~and~\eqref{u-eq}  should be with respect to conformal time defined by $a d\tau=d t$, but this makes no difference to our conclusions when expanding around $t=t_{\gamma}$ where $a \sim a(t_{\gamma}$) is nearly constant.)  The behavior of $z$ and $v$ is very similar to the case of inflation when the inflaton field approaches a turning point during oscillations about the true minimum during reheating, which is  known to be non-singular \cite{Mukhanov:1990me}.
 Since the sound speeds are ${\cal O}(1)$ at the zero-crossing (and for all $t$) in our examples, there are not any of the usual signs of a strong coupling problem.  Future work will include a complete analysis of this issue and a fully non-linear treatment using the techniques of numerical general relativity \cite{num-bounce}.

Remarkably, the zero-crossing for $\gamma$ can be shifted to the infinite past ($t_{\gamma}\to -\infty$) such that $\gamma$ can stay positive for arbitrarily long finite times before the bounce provided  $B$ and $B_h$ approach zero rapidly enough for $I(t)$ in Eq.~(\ref{BeqInt})  to converge to a finite value as $t \rightarrow -\infty$.
(We note that T. Kobayashi, in presenting his no-go theorem in Ref.~\cite{Kobayashi:2016xpl}, explored the possibility of $B_h\to0$ for $t \rightarrow -\infty$ in the context of Galilean Genesis but did not consider the implications for bouncing solutions.) There is, though, an important difference compared to the case where the zero-crossing occurs at a finite time $t_{\gamma}$ because   $H\to0$ and $a\to\infty$ at $t \rightarrow -\infty$ instead of approaching non-zero values.  While it is possible to choose $A_h$ such that the tensor sound speed $c_T=1$ as $t\to -\infty$, the scalar sound speed $c_S$ approaches zero. The fact that $A_h(t)$ and $B_h(t)$ must approach zero in this limit means that gravity deviates from pure Einstein even as $t_{\gamma}\to-\infty$;  for this reason,  the infinite limit is less practical for model-building and may be best considered as an academic exercise that shows the power of our singularity resolution.

\section{Examples}
\label{examples}
%
\begin{figure}
\label{fig0}
\includegraphics[width=8.5cm]{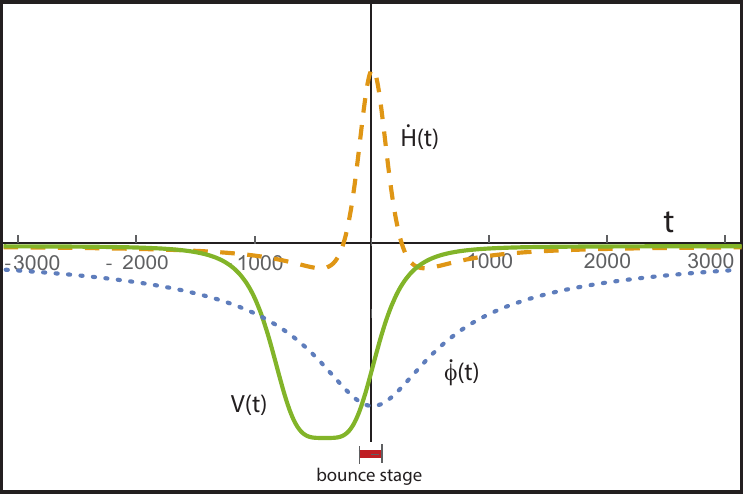}
\caption{An example of  input background solutions that can generate a stable, non-singular bounce:
 $\dot{H}$ (dashed orange curve), $\dot{\phi}(t)$ (dotted blue curve) and $V(\phi(t))\equiv V(t)$ (solid green curve); plots of $H(t)$ and $\gamma(t)$ are shown in Fig.~\ref{fig3}.  The time coordinate is given in reduced Planck units;
the units for the $y$-axis are arbitrary; the functions are rescaled for the purposes of illustrating their shapes.  The specific functional forms are given in the text.  The range around $t=0$ where $\dot{H}>0$ is the bounce stage that encompasses the period of NEC-violation.}
\label{fig0}
\end{figure}
To construct examples of fully stable cosmological solutions with a non-singular classical bounce employing the `inverse method,' we make use of the fact that, once ${\cal L}_4$ is included, we have enough degrees of freedom that we can independently choose the background behavior $H(t)$,  the behavior of $ \gamma(t)$ derived from the shift constraint, the dynamics of the tensor sector  $A_h(t)$ and $B_h(t)$, and the potential $V(t)$ and that this freedom is enough to find a broad family of solutions in which  $A(t)$ and $B(t)$ are non-negative for all $t$.  See Fig.~\ref{fig0} for a particular example corresponding to the background profile given by $H(t) = d_1 t/(1 +3  d_1 t^2)$  where $d_1= 5.5 \times 10^{-6}$; $\dot{\phi}(t) =-1/(1 + d_1 t^2)^{1/2}$;  and $V(t):= -d_2/[1+ \{d_3(t-t_{\gamma})\}^4]$ where $d_2= 4 \times 10^{-6}$, $d_3= 2.2 \times 10^{-3}$ and $t_{\gamma}=-390$.  In addition, the example uses  $A_h(t)=B_h(t)=d_5(t-t_{\gamma})^2/(1+ (d_5(t-t_{\gamma}))^2)$, where $d_5= 10^{-5}$.  The action converges to pure Einstein gravity and scale factor diverges as $t \rightarrow \pm \infty$, so the solution is geodesically complete.

Using these input functions, the couplings $f_1, f_2, b, k$, and $q$ as a function of time $t$ can be determined using the following relations:
\begin{eqnarray}
\label{f_1(t)}
f_1(t)  &=&  \frac{A_{h} + B_{h}}{2} -1 
\,,\\
\label{f_2(t)}
f_2(t) &=& \frac{B_{h}-A_{h}}{2\,\dot{\phi}^2} 
\,,
\\
\label{b(t)}
b(t) &=& \frac{2}{\dot{\phi}^3} \left(- \gamma + \left( 2A_h - B_h \right)H + \frac{1}{2}\dot{f}_1 + \frac{3}{2}\dot{f}_2\dot{\phi}^2\right)
\,,
\\
\label{EF1-FRW-k}
k(t)  &=& 
-\frac{2}{\dot{\phi}^2} \bigg( \dot{\gamma} + 3H\gamma + \frac{d}{dt}\left( \big( A_{h} + B_{h} \right)H \big)\\
\nonumber
&&\qquad + \frac{3}{2}\left( A_{h} + B_{h} \right)H^2 +\ddot{f}_1 -2V \bigg)
\,,
\\
\label{EF2-FRW-q}
q(t)
&=& \frac{4}{3\dot{\phi}^4}\bigg(\dot{\gamma}+ 9H\gamma + \frac{d}{dt}\left( \left( A_{h} + B_{h} \right)H \right)  \\
\nonumber
&&\qquad + \frac{9}{2}\left(B_{h}- A_{h}  \right)H^2 + \ddot{f}_1 -3V \bigg) + \frac{2}{3}b'
\,;
\end{eqnarray}
the couplings as a function of $\phi$ follow immediately after substituting the inverse function $t(\phi)$ for the variable $t$.
The first three relations follow from the definition of $A_h, B_h$ and $\gamma$ while the last two relations can be easily derived from the background equations~(\ref{EF1-FRW}-\ref{EF2-FRW}). We note that the inverse method is not constrained to the specific Lagrangian given by Eqs.~(\ref{g2S}-\ref{g5S}) but straightforwardly applies to arbitrary Galileon couplings $G_i(X, \phi)$.  
\begin{figure}[!t]
\includegraphics[width=8.75cm]{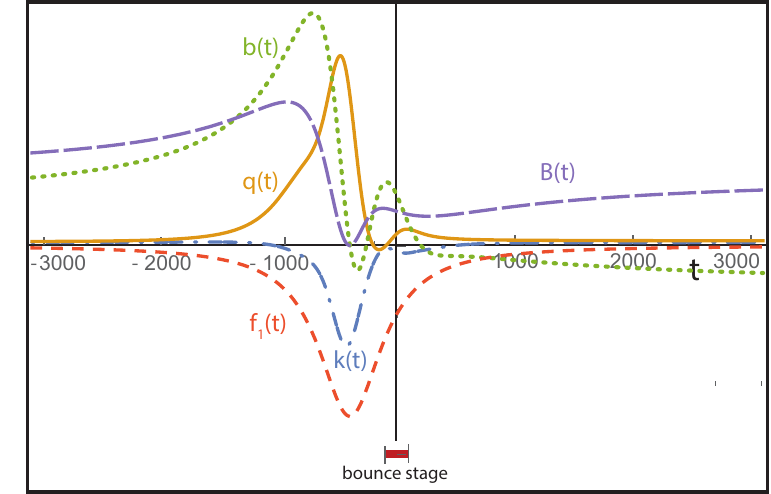}
\caption{A plot of the dimensionless kinetic coefficients in  Eqs.~(\ref{f_1(t)}-\ref{EF2-FRW-q}) obtained using the inverse method.  The time coordinate is given in Planck units; the units for the $y$-axis are arbitrary; the functions are rescaled for the purposes of illustrating their shapes. Quantitatively, the coupling parameters are several orders of magnitude  smaller than unity throughout the evolution.}
\label{fig1}
\end{figure}

Finally,  we can compute $m_{\alpha}$ as a function of the input dynamical variables $H, \dot{\phi}, A_h, B_h, \gamma$ and $V(t)$,
\begin{eqnarray}
m_{\alpha}(t) &=& \dot{\gamma} + 3H\gamma  +  \frac{d^2}{dt^2} \left( \frac{A_{h} + B_{h}}{2} \right) + 3H\frac{d}{dt} \left( \frac{A_{h} + B_{h}}{2} \right) \nonumber
\\
&+& 3\left(A_{h} + B_{h} \right)H^2  + \frac{d}{dt}\left( \left( A_{h} + B_{h} \right)H \right)  - 4V
,\quad
\end{eqnarray}
substitute into  Eq.~\eqref{Aeq} and check that $A(t)$ is non-negative.  As noted above, the freedom to introduce a potential makes this condition easy to attain.  
In our example, for the purposes of illustration, we selected $V(\phi)$ corresponding to a typical ekpyrotic potential \cite{Steinhardt:2001st}; see Fig.~\ref{fig0}.
For $A_h=B_h=1$, Eqs.~(\ref{f_1(t)}-\ref{EF2-FRW-q}) reduce to the corresponding expressions for the generalized cubic Galileon that we derived in Ref.~\cite{Ijjas:2016tpn}.

\begin{figure}[!t]
\includegraphics[width=8.25cm]{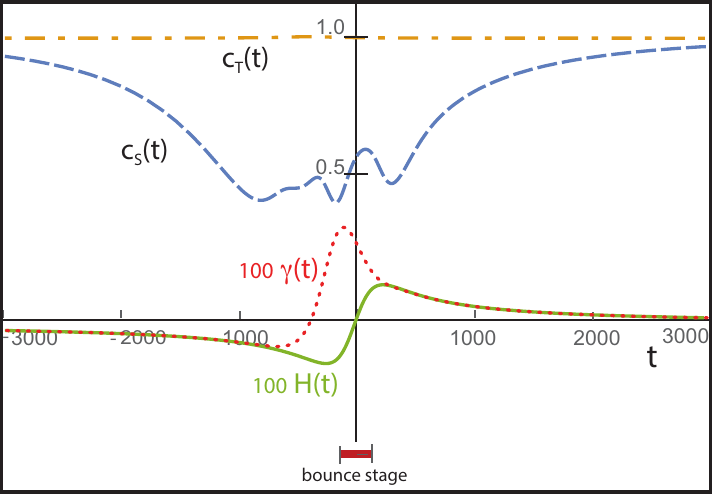}
\caption{A plot of the sound speed $c_S\equiv (B(t)/A(t))^{1/2}$ (dashed blue curve)  for co-moving curvature modes and $c_T\equiv (B_h(t)/A_h(t))^{1/2}$ for tensor modes as a function of time $t$ where $\gamma(t)$ (dotted red curve) passes through zero at $t_{\gamma}<0$, where $t_{\gamma}$ is during the NEC-satisfying stage prior to the bounce stage.   
$c_S(t)$  approaches unity and  $\gamma(t)$ approaches $H(t)$ (green solid curve) as $t \rightarrow \pm \infty$, which corresponds to approaching a contracting (expanding) universe described by Einstein gravity in the two limits; $c_T=1$ throughout.    The units for the $y$-axis have  reduced Planck mass equal to unity where appropriate; $\gamma(t)$ and $H(t)$ are rescaled as indicated for the purposes of illustrating their shapes on the same plot.}
\label{fig3}
\end{figure}

Employing the inverse method, the analysis of the linear theory dramatically simplifies. 
 It is now possible to systematically and rapidly search for fully stable, non-singular bouncing background solutions that show the asymptotic behavior outlined in the previous section.  
Using the example in Figure~\ref{fig0}, the resulting kinetic coefficients are shown in Fig.~\ref{fig1}.
In Figure~\ref{fig3}, we have plotted the corresponding sound speeds. For all times, the tensor sound speed is constant, $c_T^2\equiv1$ and the sound speed for co-moving curvature modes is real ($A(t),\, B(t)>0$) and subluminal, with $0\ll c_S^2 \sim {\cal O}(1)$.  In the asymptotic past and future, $c_S^2 \to 1$, and the theory approaches pure Einstein gravity.

We emphasize that, in applying the inverse method, one must specify more than the desired background solutions $H(t)$ and $\phi(t)$ as input since they do not form a complete set for characterizing the linear theory. Rather, it is important to choose the right combination of dynamical background variables that fully describe the linear theory, as exemplified above.

\section{Discussion}

In this Letter, we presented geodesically complete, stable non-singular bouncing cosmological solutions that are non-pathological over all time, confirming and extending our previous result in Ref.~\cite{Ijjas:2016tpn} where we demonstrated that it is possible to construct solutions that are non-pathological during the NEC-violating bounce stage.  
The key was to identify the source of the singular behavior in cubic Galileon cosmologies, {\it i.e.,} actions where the coefficients of ${\cal L}_4$ and ${\cal L}_5$ are set to zero.  What we have shown is that the bad behavior in this case is not directly related to the NEC-violating bounce stage, but to the fact that the Hubble parameter $H(t)$ switches sign at some point during cosmic evolution.  Before or after $H(t)$ changes sign, the dynamical quantity derived from the shift constraint, $\gamma(t)$, has to change sign as well, which is what causes the pathological behavior.  In fact, the pathological behavior arises in cubic Galileon cosmologies that smoothly transit from expansion to contraction without any bounce or NEC violation. This 
observation explains why earlier authors were able to find non-pathological NEC-violating solutions in cubic Galileon genesis models where $H(t)$ does not change sign ({\it e.g.}, see \cite{Pirtskhalava:2014esa}), but failed to find fully stable solutions with a bounce.

Notably, the pathology of the cubic Galileon action is resolved in a natural way, simply by extending the  action to include the next-order ${\cal L}_4$ interaction.  Using this extension, we showed that we can construct geodesically complete, stable, non-singular bounce solutions.  Another natural extension is to include additional degrees of freedom corresponding to NEC-satisfying matter and radiation \cite{Ijjas1}. The resulting construction of a fully stable bouncing solution should give one pause.  
Together with Ref.~\cite{Ijjas:2016tpn}, it removes the last major roadblock that has been holding back interest in cosmologies that explain the origin of the large-scale structure of the universe in terms of a contracting phase connecting to the current expanding phase through a cosmological bounce.

{\it Acknowledgements.} We thank Frans Pretorius and Vasileios Paschalidis  for helpful discussions. This research was partially supported by the U.S. Department of Energy under grant number DEFG02-91ER40671.

\bibliographystyle{apsrev}
\bibliography{simple-exit}

\begin{thebibliography}{16}
\expandafter\ifx\csname natexlab\endcsname\relax\def\natexlab#1{#1}\fi
\expandafter\ifx\csname bibnamefont\endcsname\relax
  \def\bibnamefont#1{#1}\fi
\expandafter\ifx\csname bibfnamefont\endcsname\relax
  \def\bibfnamefont#1{#1}\fi
\expandafter\ifx\csname citenamefont\endcsname\relax
  \def\citenamefont#1{#1}\fi
\expandafter\ifx\csname url\endcsname\relax
  \def\url#1{\texttt{#1}}\fi
\expandafter\ifx\csname urlprefix\endcsname\relax\def\urlprefix{URL }\fi
\providecommand{\bibinfo}[2]{#2}
\providecommand{\eprint}[2][]{\url{#2}}

\bibitem[{\citenamefont{Khoury et~al.}(2001)\citenamefont{Khoury, Ovrut,
  Steinhardt, and Turok}}]{Khoury:2001wf}
\bibinfo{author}{\bibfnamefont{J.}~\bibnamefont{Khoury}},
  \bibinfo{author}{\bibfnamefont{B.~A.} \bibnamefont{Ovrut}},
  \bibinfo{author}{\bibfnamefont{P.~J.} \bibnamefont{Steinhardt}},
  \bibnamefont{and} \bibinfo{author}{\bibfnamefont{N.}~\bibnamefont{Turok}},
  \bibinfo{journal}{Phys. Rev.} \textbf{\bibinfo{volume}{D64}},
  \bibinfo{pages}{123522} (\bibinfo{year}{2001}), \eprint{hep-th/0103239}.

\bibitem[{\citenamefont{Lehners et~al.}(2007)\citenamefont{Lehners, McFadden,
  Turok, and Steinhardt}}]{Lehners:2007ac}
\bibinfo{author}{\bibfnamefont{J.-L.} \bibnamefont{Lehners}},
  \bibinfo{author}{\bibfnamefont{P.}~\bibnamefont{McFadden}},
  \bibinfo{author}{\bibfnamefont{N.}~\bibnamefont{Turok}}, \bibnamefont{and}
  \bibinfo{author}{\bibfnamefont{P.~J.} \bibnamefont{Steinhardt}},
  \bibinfo{journal}{Phys.Rev.} \textbf{\bibinfo{volume}{D76}},
  \bibinfo{pages}{103501} (\bibinfo{year}{2007}), \eprint{hep-th/0702153}.

\bibitem[{\citenamefont{Buchbinder et~al.}(2007)\citenamefont{Buchbinder,
  Khoury, and Ovrut}}]{Buchbinder:2007ad}
\bibinfo{author}{\bibfnamefont{E.~I.} \bibnamefont{Buchbinder}},
  \bibinfo{author}{\bibfnamefont{J.}~\bibnamefont{Khoury}}, \bibnamefont{and}
  \bibinfo{author}{\bibfnamefont{B.~A.} \bibnamefont{Ovrut}},
  \bibinfo{journal}{Phys.Rev.} \textbf{\bibinfo{volume}{D76}},
  \bibinfo{pages}{123503} (\bibinfo{year}{2007}), \eprint{hep-th/0702154}.

\bibitem[{\citenamefont{Gielen and Turok}(2016)}]{Gielen:2015uaa}
\bibinfo{author}{\bibfnamefont{S.}~\bibnamefont{Gielen}} \bibnamefont{and}
  \bibinfo{author}{\bibfnamefont{N.}~\bibnamefont{Turok}},
  \bibinfo{journal}{Phys. Rev. Lett.} \textbf{\bibinfo{volume}{117}},
  \bibinfo{pages}{021301} (\bibinfo{year}{2016}), \eprint{1510.00699}.

\bibitem[{\citenamefont{Bars et~al.}(2013)\citenamefont{Bars, Steinhardt, and
  Turok}}]{Bars:2013vba}
\bibinfo{author}{\bibfnamefont{I.}~\bibnamefont{Bars}},
  \bibinfo{author}{\bibfnamefont{P.~J.} \bibnamefont{Steinhardt}},
  \bibnamefont{and} \bibinfo{author}{\bibfnamefont{N.}~\bibnamefont{Turok}},
  \bibinfo{journal}{Phys.Lett.} \textbf{\bibinfo{volume}{B726}},
  \bibinfo{pages}{50} (\bibinfo{year}{2013}), \eprint{1307.8106}.

\bibitem[{\citenamefont{Ijjas and Steinhardt}(2016)}]{Ijjas:2016tpn}
\bibinfo{author}{\bibfnamefont{A.}~\bibnamefont{Ijjas}} \bibnamefont{and}
  \bibinfo{author}{\bibfnamefont{P.~J.} \bibnamefont{Steinhardt}},
  \bibinfo{journal}{Phys. Rev. Lett.} \textbf{\bibinfo{volume}{117}},
  \bibinfo{pages}{121304} (\bibinfo{year}{2016}), \eprint{1606.08880}.

\bibitem[{\citenamefont{Horndeski}(1974)}]{Horndeski:1974wa}
\bibinfo{author}{\bibfnamefont{G.~W.} \bibnamefont{Horndeski}},
  \bibinfo{journal}{Int. J. Theor. Phys.} \textbf{\bibinfo{volume}{10}},
  \bibinfo{pages}{363} (\bibinfo{year}{1974}).

\bibitem[{\citenamefont{de~Rham}(2012)}]{deRham:2012az}
\bibinfo{author}{\bibfnamefont{C.}~\bibnamefont{de~Rham}},
  \bibinfo{journal}{Comptes Rendus Physique} \textbf{\bibinfo{volume}{13}},
  \bibinfo{pages}{666} (\bibinfo{year}{2012}), \eprint{1204.5492}.

\bibitem[{\citenamefont{Khoury et~al.}(2011)\citenamefont{Khoury, Lehners, and
  Ovrut}}]{Khoury:2011da}
\bibinfo{author}{\bibfnamefont{J.}~\bibnamefont{Khoury}},
  \bibinfo{author}{\bibfnamefont{J.-L.} \bibnamefont{Lehners}},
  \bibnamefont{and} \bibinfo{author}{\bibfnamefont{B.~A.} \bibnamefont{Ovrut}},
  \bibinfo{journal}{Phys.Rev.} \textbf{\bibinfo{volume}{D84}},
  \bibinfo{pages}{043521} (\bibinfo{year}{2011}), \eprint{1103.0003}.

\bibitem[{\citenamefont{Kobayashi et~al.}(2011)\citenamefont{Kobayashi,
  Yamaguchi, and Yokoyama}}]{Kobayashi:2011nu}
\bibinfo{author}{\bibfnamefont{T.}~\bibnamefont{Kobayashi}},
  \bibinfo{author}{\bibfnamefont{M.}~\bibnamefont{Yamaguchi}},
  \bibnamefont{and} \bibinfo{author}{\bibfnamefont{J.}~\bibnamefont{Yokoyama}},
  \bibinfo{journal}{Prog. Theor. Phys.} \textbf{\bibinfo{volume}{126}},
  \bibinfo{pages}{511} (\bibinfo{year}{2011}), \eprint{1105.5723}.

\bibitem[{\citenamefont{Mukhanov et~al.}(1992)\citenamefont{Mukhanov, Feldman,
  and Brandenberger}}]{Mukhanov:1990me}
\bibinfo{author}{\bibfnamefont{V.~F.} \bibnamefont{Mukhanov}},
  \bibinfo{author}{\bibfnamefont{H.}~\bibnamefont{Feldman}}, \bibnamefont{and}
  \bibinfo{author}{\bibfnamefont{R.~H.} \bibnamefont{Brandenberger}},
  \bibinfo{journal}{Phys.Rept.} \textbf{\bibinfo{volume}{215}},
  \bibinfo{pages}{203} (\bibinfo{year}{1992}).

\bibitem[{\citenamefont{Ijjas et~al.}(2017)\citenamefont{Ijjas, Paschalidis,
  Pretorius, and Steinhardt}}]{num-bounce}
\bibinfo{author}{\bibfnamefont{A.}~\bibnamefont{Ijjas}},
  \bibinfo{author}{\bibfnamefont{V.}~\bibnamefont{Paschalidis}},
  \bibinfo{author}{\bibfnamefont{F.}~\bibnamefont{Pretorius}},
  \bibnamefont{and} \bibinfo{author}{\bibfnamefont{P.~J.}
  \bibnamefont{Steinhardt}}, \bibinfo{journal}{{in preparation}}
  (\bibinfo{year}{2017}).

\bibitem[{\citenamefont{Kobayashi}(2016)}]{Kobayashi:2016xpl}
\bibinfo{author}{\bibfnamefont{T.}~\bibnamefont{Kobayashi}},
  \bibinfo{journal}{Phys. Rev.} \textbf{\bibinfo{volume}{D94}},
  \bibinfo{pages}{043511} (\bibinfo{year}{2016}), \eprint{1606.05831}.

\bibitem[{\citenamefont{Steinhardt and Turok}(2002)}]{Steinhardt:2001st}
\bibinfo{author}{\bibfnamefont{P.~J.} \bibnamefont{Steinhardt}}
  \bibnamefont{and} \bibinfo{author}{\bibfnamefont{N.}~\bibnamefont{Turok}},
  \bibinfo{journal}{Phys.Rev.} \textbf{\bibinfo{volume}{D65}},
  \bibinfo{pages}{126003} (\bibinfo{year}{2002}), \eprint{hep-th/0111098}.

\bibitem[{\citenamefont{Pirtskhalava et~al.}(2014)\citenamefont{Pirtskhalava,
  Santoni, Trincherini, and Uttayarat}}]{Pirtskhalava:2014esa}
\bibinfo{author}{\bibfnamefont{D.}~\bibnamefont{Pirtskhalava}},
  \bibinfo{author}{\bibfnamefont{L.}~\bibnamefont{Santoni}},
  \bibinfo{author}{\bibfnamefont{E.}~\bibnamefont{Trincherini}},
  \bibnamefont{and}
  \bibinfo{author}{\bibfnamefont{P.}~\bibnamefont{Uttayarat}},
  \bibinfo{journal}{JHEP} \textbf{\bibinfo{volume}{12}}, \bibinfo{pages}{151}
  (\bibinfo{year}{2014}).

\bibitem[{\citenamefont{Ijjas}(to appear)}]{Ijjas1}
\bibinfo{author}{\bibfnamefont{A.}~\bibnamefont{Ijjas}} (\bibinfo{year}{to
  appear}).

\end{thebibliography}

\end{document}